# Training Future Engineers to Be Ghostbusters: Hunting for the Spectral Environmental Radioactivity

**Matteo Albéri** [1,2], *****, **Marica Baldoncini** [1,2], **Carlo Bottardi** [1,3], **Enrico Chiarelli** [1,2],
**Sheldon Landsberger** [4], **Kassandra Giulia Cristina Raptis** [1,2], **Andrea Serafini** [1,3], **Virginia Strati** [1,3] and
**Fabio Mantovani** [1,3]

1   Department of Physics and Earth Sciences, University of Ferrara, Via Saragat 1, 44121, Ferrara, Italy;
    baldoncini@fe.infn.it (M.B.); bottardi@fe.infn.it (C.B.); enrico.chiarelli@student.unife.it (E.C);
    kassandra.raptis@lnl.infn.it (K.G.C.R); serafini@fe.infn.it (A.S.); strati@fe.infn.it (V.S.); mantovani@fe.infn.it (F.M.)
2   INFN, Legnaro National Laboratories, Viale dell'Università, 2, 35020, Legnaro, Padua, Italy
3   INFN, Ferrara Section, Via Saragat 1, 44121, Ferrara, Italy
4   Nuclear Engineering Teaching Lab, University of Texas, Pickle Research Campus R-9000, Austin, TX 78712, USA;
    s.landsberger@mail.utexas.edu (S.L.)

*   Correspondence: alberi@fe.infn.it; Tel.: +39-329-0715-328

**Abstract:** Although environmental radioactivity is all around us, the collective public imagination often associates a negative feeling to this natural phenomenon. To increase the familiarity with this phenomenon we have designed, implemented, and tested an interdisciplinary educational activity for pre-collegiate students in which nuclear engineering and computer science are ancillary to the comprehension of basic physics concepts. Teaching and training experiences are performed by using a 4" × 4" NaI(Tl) detector for in-situ and laboratory γ-ray spectroscopy measurements. Students are asked to directly assemble the experimental setup and to manage the data-taking with a dedicated Android app, which exploits a client-server system that is based on the Bluetooth communication protocol. The acquired γ-ray spectra and the experimental results are analyzed using a multiple-platform software environment and they are finally shared on an open access Web-GIS service. These all-round activities combining theoretical background, hands-on setup operations, data analysis, and critical synthesis of the results were demonstrated to be effective in increasing students' awareness in quantitatively investigating environmental radioactivity. Supporting information to the basic physics concepts provided in this article can be found at http://www.fe.infn.it/radioactivity/educational.

# 1. Introduction

In the last decade, various educational approaches have been developed by different scientists and teachers with the aim of giving a clear picture about radiation issues [1,2]. In the public domain, radioactivity can evoke negative feelings that are associated to nuclear accidents or radioactive waste management or diseases [3,4]. In this perspective, one of the missions of traditional radiation physics lectures is to make students aware that radiation from the ground and from the sky is all around us, with much of it passing through us constantly and that even food and our bodies are radioactive, to a degree. In this paper, we present two educational activities for pre-collegiate students, which adopt a mixed method that is based on applying nuclear engineering concepts and computer science tools to explore in-situ environmental radioactivity.

The two teaching and training activities are addressed to a group of 5–8 students and they are conceived as 4-h hands-on experiments (Table 1) involving the use of multiplatform software (Android, Windows) for the gamma spectra acquisition and analysis and for ad-hoc Web-GIS applications. A 4" × 4" thallium-activated sodium iodide (NaI(Tl)) scintillation detector, a relatively accessible and affordable instrument is employed in both experiments. Its high detection efficiency and the fact that it works at room temperature given the possibility of performing quick and reliable measurements in different experimental conditions, as typically requested in the case of educational experiences [5-7]. The experimental setup also consists of wireless dedicated nuclear electronics for the digitization of the signal, which integrates data storage as well as a data-taking programming capability in terms of main experimental parameters (acquisition time, operating voltage, etc.).

The indoor experiment is designed as a propaedeutic experience for the comprehension of the nature of gamma photons and of the main features characterizing a gamma-ray spectrum acquired with a scintillation detector. The outdoor experiment is structured in multiple in-situ measurements over different ground coverage types. This activity is intended to familiarize the students with the range of radioactivity levels that are present in the environment and to aid their critical understanding of the measurement of spatial resolution and of the spatial distribution of radioactivity in the investigated area.

This educational path was tested preliminarily, involving pre-collegiate students and teachers of Italian high schools and improved in the framework of the Maymester "Concepts in Nuclear and Radiation Engineering" developed from an international cooperation between the University of Ferrara (Italy) and the Cockrell School of Engineering at the University of Texas at Austin (USA).

**Table 1.** The supplies (equipment and software) used for the two experiments carried out during the educational activities, together with the corresponding educational aims. In both cases, the measurements are performed with a 4" × 4" thallium-activated sodium iodide (NaI(Tl)) detector and a MultiChannel Analyzer (MCA) γstream by CAEN.

| | Equipment and Software | Educational Aims |
|---|---|---|
| Indoor experiment | • 4" × 4" NaI(Tl) detector with MultiChannel Analyzer (MCA) γstream by CAEN<br>• Tablet or smartphone<br>• GammaTOUCH app<br>• Lead slabs<br>• Point-like radioactive source ($^{137}$Cs)<br>• Aluminum layers | • Assembling an experimental setup critically understanding the functioning of the components and the operation mode<br>• Learning how to interpret a γ-ray spectrum acquired with a scintillation detector<br>• Using an ad hoc Android app for gamma spectra analysis<br>• Understanding of the high penetration capacity of gamma photons<br>• Determination of linear attenuation coefficients of gamma radiation |
| Outdoor experiment | • 4" × 4" NaI(Tl) detector with MultiChannel Analyzer (MCA) γstream by CAEN<br>• Tablet or smartphone<br>• GammaTOUCH app<br>• Google Maps<br>• Google Earth | • Learning how to design and perform in-situ γ-ray spectrometry measurements<br>• Exploiting the potentialities of a client-server system based on a Bluetooth communication protocol<br>• Critical understanding of all unplanned factors making an in-situ γ-ray survey a complex issue<br>• Adopting an open access Web-GIS platform to share and visualize the results through an interactive GUI |

## 2. Theoretical Background

Radioactivity is a physical phenomenon occurring when an unstable nucleus undergoes a transition from one energy state to another and it is typically measured in becquerels, corresponding to one decay per second. Natural or artificial radiation sources can be found everywhere, starting from the first moments of life of our universe. Natural sources include the cosmogenic radionuclides, which are related to the interaction between cosmic rays and nuclei of atoms in the atmosphere, and the so-called primordial radionuclides existing since the Earth formed and that have not completely decayed due to their long half-life (~$10^9$ yr and longer). The most common isotopes in the Earth responsible for the so called terrestrial radiation are $^{238}$U, $^{232}$Th, and $^{40}$K, together with their multiple daughter products. Although $^{235}$U is also present, it is not considered as its isotopic abundance is 0.72%, to be compared with the 99.28% isotopic abundance of $^{238}$U. It is estimated that 80% of the average annual dose for the world's population comes from natural background radiation [8].

While $^{40}$K undergoes one single decay, $^{238}$U and $^{232}$Th produce decay chains that comprise α, β, and/or γ decays. The γ decays, in contrast to α and β, do not change the atomic number of the nuclei: they occur when a nucleus in an excited state, which is often produced by a previous decay (typically alpha or beta), emits a photon, called a γ ray, in order to reach a more stable configuration [9]. γ rays have the same physical nature as visible light but belong to a region of the electromagnetic spectrum characterized by higher frequencies (i.e., higher energies, tens to thousands of keV): as a consequence, they are invisible to our eyes and a detector is needed in order to reveal them.

Uranium-238 has a half-life of $4.47 \times 10^9$ years and its decay chain comprises 18 unstable isotopes among which the main gamma emitters are $^{234m}$Pa, $^{214}$Pb, and $^{214}$Bi. $^{232}$Th has a half-life of $1.41 \times 10^{10}$ years, its decay chain includes 12 unstable isotopes, among which the main gamma emitters are $^{228}$Ac, $^{212}$Pb, $^{212}$Bi, and $^{208}$Tl. Here, we measure $^{214}$Bi and $^{208}$Tl by monitoring gamma lines having an energy of 1764 keV and 2614 keV, respectively. In particular, the 2614 keV gamma emission from $^{208}$Tl corresponds to the endpoint of the terrestrial γ-ray spectrum that is associated with the $^{232}$Th decay chain. Argon-40, the daughter of $^{40}$K decays, produces a gamma signal at 1460 keV, which is usually a distinctive feature of the environmental gamma spectrum.

A photon can interact with matter mainly via three processes: the photoelectric effect, the Compton scattering, and the pair production-annihilation [10]. Through these phenomena, the energy of the γ rays is deposited in a given material in the form of kinetic energy of electrons. The photoelectric effect is predominant for low energies and it arises from the absorption of a photon by an atom and the ejection of an electron from one of the atomic bound shells. Compton scattering is the main interaction mechanism of terrestrial gamma photons, as it dominates at intermediate energies (~1 MeV). It is the process by which a photon scatters from a nearly free atomic electron, resulting in a less energetic photon and a scattered electron carrying the energy lost by the photon. The energy that is gathered by the scattered electron (and finally deposited in the detector) is a continuous function of the scattering angle and it is what gives rise in a measured spectrum to the Compton continuum (Figure 1b). The maximum energy transferable to the electron in a single collision is obtained for backscattered photon and it is at the origin of the formation of the so called Compton edge (Figure 1b). The pair production process corresponds to the conversion of a γ ray into an electron-positron pair and it occurs only if the gamma has a minimum energy equal to the mass of the particle pair (2 $m_e c^2$ = 1022 keV).

If one considers a gamma beam propagating in matter in the direction $x$ and if $N_0$ corresponds to the initial number of gammas, due to the interplay of the three attenuating processes described before, the beam loses a number of photons $\Delta N$. The relative photon loss $\Delta N/N$ is proportional to the covered distance $\Delta x$:

$$\frac{\Delta N}{N} = -\mu_{mass} \rho \Delta x, \qquad (1)$$

with $\mu_{mass}$ in cm$^2$/g being the mass attenuation coefficient of the traversed material [11]. Figure 1a shows the overall $\mu_{mass}$ for aluminum as a function of the gamma energy, together with the relative contributions of the three interaction mechanisms. The linear attenuation coefficient $\mu$ represents the inverse of the distance at which the number of photons is reduced by a factor $1/e$, as can be inferred by the following equation [12]:

$$N = N_0 e^{-\mu x}, \qquad (2)$$

where $\mu$ in cm$^{-1}$ is obtained by multiplying $\mu_{mass}$ times the material density $\rho$ in g/cm$^3$. In order to experimentally test this theory, a monoenergetic photon beam can be produced by surrounding a point-like radioactive source that is characterized by a single gamma emission with a lead box having a small hole on one side, which is meant to produce a collimating effect.

The experiments are performed by using a 4" × 4" (NaI(Tl)) detector that can acquire a γ-ray spectrum (see Section 3), i.e., a histogram of events classified according to the energy deposited inside the detector itself. A typical feature that can be observed in a γ-ray spectrum is the so-called photopeak, which is populated by those events in which a gamma photon, having energy that is equal to the one of the decay, impinges on the scintillator depositing its full energy in the active detector volume (Figure 1b).

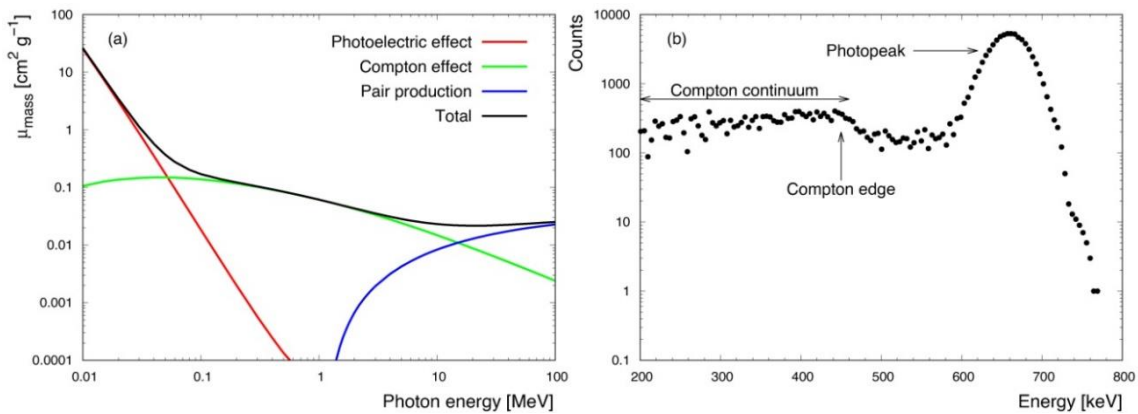

**Figure 1.** (**a**) Mass attenuation coefficient ($\mu_{mass}$) for aluminum as function of the photon energy (source: https://physics.nist.gov): the three contributions due to the photoelectric, Compton and pair production interactions are separately displayed. (**b**) Example of a gamma spectrum acquired by juxtaposing a $^{137}$Cs point-like source to a non-shielded NaI(Tl) detector in which the photopeak shape centered at 661.7 keV is clearly visible, together with the lower energy Compton continuum and the structure of the Compton edge.

Equation (2) is the key for understanding the lateral horizon of in-situ γ-ray spectroscopy. The horizontal field of view of a γ-ray detector expresses the relative contribution to the total signal that is produced within a given radial distance from the detector vertical axis. The lateral horizon depends on the height of the detector: for instance, a spectrometer that was placed at ground level receives 90% of the signal from a radius of ~0.5 m (Figure 2); at a height of 0.5 m, 90% of the signal come from a radius of ~8 m (Figure 2 of [11]).

Supporting information to the basic concepts that are provided in this section can be found at http://www.fe.infn.it/radioactivity/educational, a website designed for teachers and students who want to deal with the topic of environmental gamma radioactivity using the didactic approach that is described in this paper.

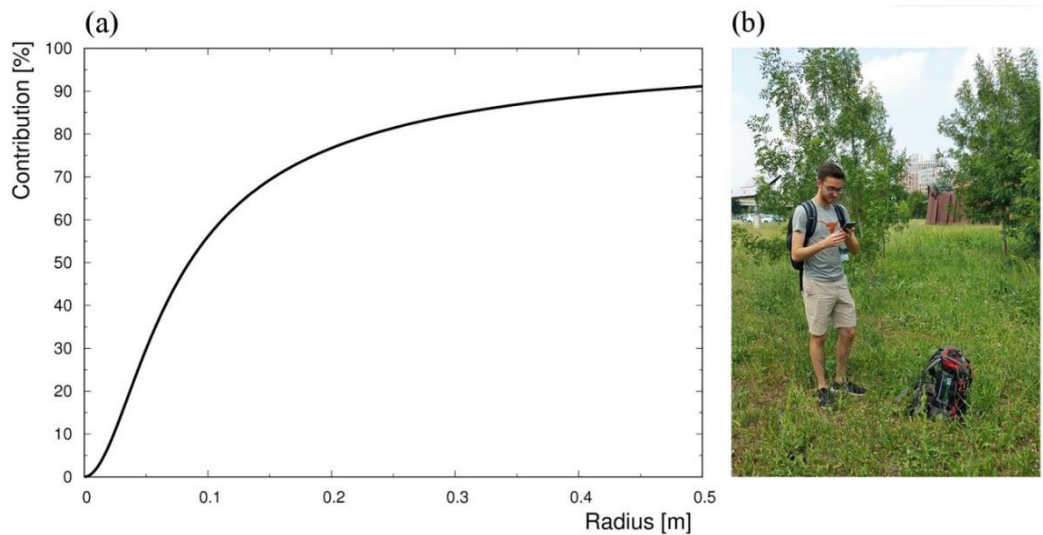

**Figure 2.** (**a**) Cumulative percentage contribution to the 1460 keV ($^{40}$K) unscattered γ signal as function of the radial distance from the vertical symmetry axis of the detector placed at 5 cm above the ground, assuming a homogeneous radioactive content of the soil. (**b**) A student performing an in-situ measurement with the backpack placed on the ground.

## 3. Indoor Experiment

The indoor experiment functions as preparatory training for the outdoor experiment depicted in Section 4. By assembling an experimental setup in the laboratory, students learn about the mode of operation of a detector during an environmental γ-ray spectroscopy measurement. Using an ad-hoc Android app, the students learn to handle the measurements and visualize the acquired spectra. Finally, by retrieving the counting statistics information, students are able to determine the linear attenuation coefficient of a given material [13]. This educational experience has the aim of enhancing the knowledge about radioactivity in terms of both natural and artificial sources as well as making the students gain experience of the high level of penetration of γ-rays in matter, which makes γ-ray spectroscopy an effective in-situ monitoring technique.

The experimental setup consists in a 4" × 4" NaI(Tl) detector, a PhotoMultiplier Tube (PMT) and a digital MultiChannel Analyzer (MCA, γstream by CAEN) (Figure 3).

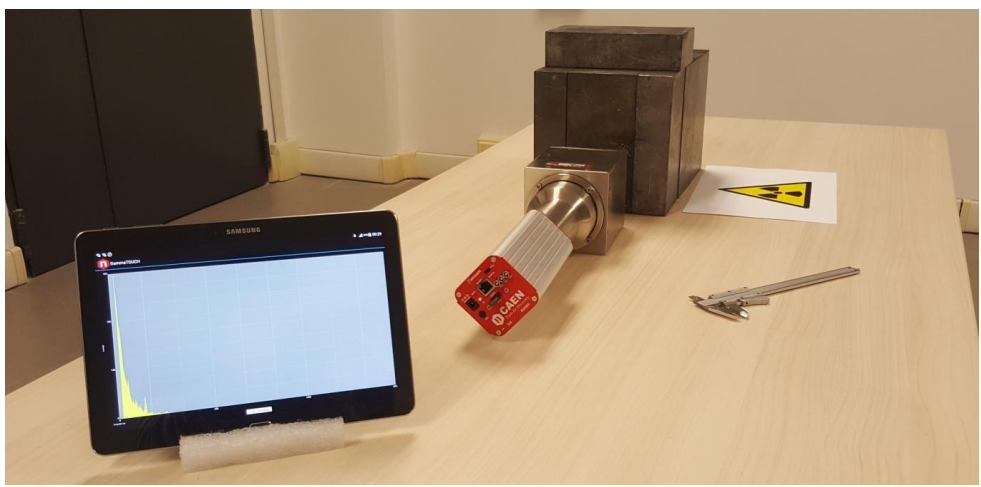

**Figure 3.** Experimental setup for the indoor experiment: on the left the tablet showing the graphical interface of the GammaTOUCH app during the acquisition, on the right the 4" × 4" NaI(Tl) detector and the lead box.

During an acquisition, the detector produces an amount of scintillation light proportional to the energy that is deposited in the NaI(Tl) crystal by the incident γ-ray. By coupling the NaI(Tl) to a PMT, scintillation light is converted to an amplified electric pulse that is proportional to the gamma energy. This signal is in turn digitized by a MCA, allowing for one to obtain recorded events classified according to the deposited energy and therefore to populate an energy spectrum. Since the NaI(Tl) crystals are characterized by a relatively high scintillation efficiency, the detector is usually able to collect sufficient statistics in a short time, also yielding an energy resolution of a level to enable radionuclide identification. The γstream can be operated via the Android app GammaTOUCH, which uses the Bluetooth communication protocol: this app allows the user to set the operating voltage, specify the acquisition time, and start the measurement. The measurements of this experiment are performed while using a collimated point-like $^{137}$Cs source that emits monochromatic gamma photons at 662 keV.

*3.1. Energy Calibration of the Gamma Spectrum*

In the first part of the experiment, students start an acquisition, setting the γstream operating voltage to 850 V and the acquisition time to 800 s. The graphic interface of the GammaTOUCH app continuously updates the histogram shape by showing the cumulative number of events over time. This preliminary step helps the students to identify the main photopeaks and to distinguish them from local fluctuations, as well as to start decrypting the information that was encoded in the different energy ranges. When the acquisition ends, the spectrum is saved in an ASCII file that lists, in a single column, the number of events for each channel and that can be opened and manipulated in an Excel spreadsheet.

A dedicated Android app performs the energy calibration of the spectrum (Figure 4), which converts the acquisition channels into energy deposited inside the detector according to the following equation:

$$E = m \cdot ch + q, \qquad (3)$$

where $E$ is the energy in keV corresponding to the channel $ch$, $m$ is the gain in keV/ch (i.e., the width of a single acquisition channel), and $q$ is the intercept in keV, corresponding to the energy of the first channel (Figure 4b).

The energy calibration procedure is based on the reconstruction of the Gaussian shapes of the $^{40}$K and $^{208}$Tl photopeaks corresponding, respectively, to the 1460 keV and 2614 keV gamma emissions. Knowing the energies of the gamma emission and the channels corresponding to the Gaussian means, the slope, and intercept of the linear relation given by (3) are calculated, as shown in the app graphical user interface (Figure 4a) and subsequently used by the students to integrate the counts of measured spectra in the energy windows of interest (Figure 5).

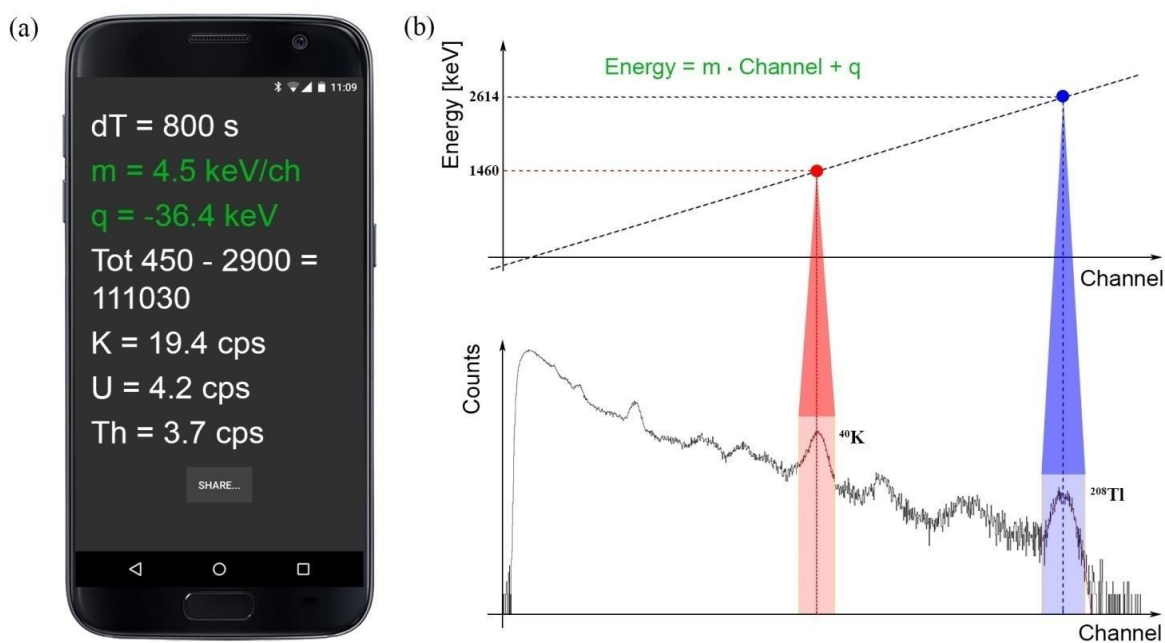

**Figure 4.** (**a**) Screenshot of the output of the Android app performing the energy calibration of the gamma spectrum: *m* and *q* represent respectively the slope and the intercept of the linear function determined on the base of the Gaussian reconstruction of the $^{40}$K and $^{208}$Tl photopeaks at 1460 keV and 2614 keV (**b**).

The integrated number of occurrences $N$ for each of the four energy windows of interest [14] (Figure 5) is obtained by summing the number of counts $N_i$ recorded in all of the energy channels belonging to the specific window. The count rate $n$ is then directly calculated by normalizing for the acquisition time $T$ as $n = \frac{N}{T}$. At the end of this procedure, the students measured the count rate in the $^{137}$Cs photopeak energy window in the absence of any $^{137}$Cs source. This measurement represents the environmental background acquisition determining the background count rate $n_{Cs\text{-}bkg}$.

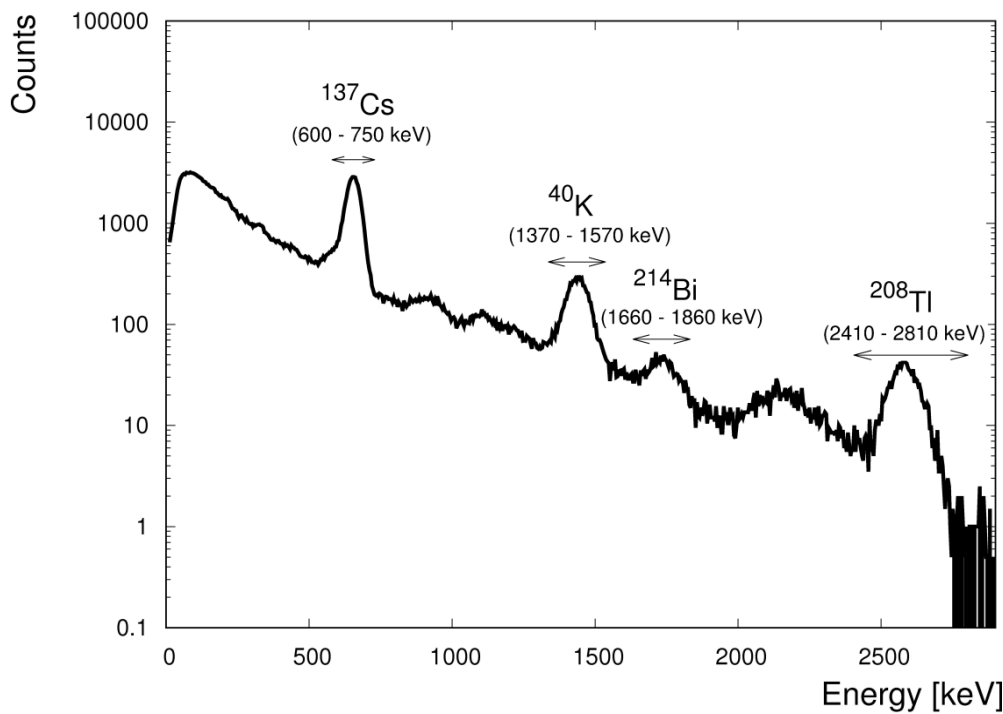

**Figure 5.** A γ-ray spectrum acquired in laboratory with a 4" × 4" NaI(Tl) detector in presence of a $^{137}$Cs point-like source for an acquisition time of 360 s. The most prominent photopeaks of $^{137}$Cs, $^{40}$K, $^{214}$Bi ($^{238}$U daughter) and $^{208}$Tl ($^{232}$Th daughter) are highlighted, together with the energy windows of interest in keV used for the count rate integration.

*3.2. Linear Attenuation Coefficient Derivation*

The goal of the second part of the experiment is to derive the linear attenuation coefficient of aluminum by using the $^{137}$Cs point-like source and 16 aluminum layers, each one having a thickness of 2.9 mm. The source is located inside a lead shielding box at a distance of 46.4 mm from the crystal bottom, which corresponds to the full thickness of all the aluminum attenuating layers (Figure 6a). The lead box has a 0.5 cm diameter hole, which allows the collimation of the gamma radiation in order to reproduce the mono-directional boundary condition described in Section 2 (see Equations (1) and (2)).

Once the experimental setup is ready, the first 360 s acquisition is started in the absence of any attenuating layer. Then, the eight successive 360 s acquisitions are performed by adding each time two aluminum layers. Finally in the last measurement, the space between the detector and the source is completely filled (Figure 6a).

The net count rate in the $^{137}$Cs photopeak $n$ is obtained by subtracting to the total count rate $n_{total}$, measured in presence of a given aluminum thickness, the background count rate $n_{Cs\text{-}bkg}$, measured during the environmental background acquisition in the first part of the experiment. The net count rate that was measured in the absence of any attenuating material $n_0$ is used as the normalization factor for each measurement performed in the presence of aluminum layers. Indeed, for each configuration, the ratio $\frac{n}{n_0}$ is plotted versus the layer thickness (Figure 6b). The Excel fitting tool is employed in order to reconstruct the exponential trend of the ratio $\frac{n}{n_0}$ as a function of the total thickness of aluminum layers, retrieving the value of the aluminum linear attenuation coefficient $\mu$ (as in Equation (2)).

The obtained result (0.184 cm$^{-1}$) is converted into a mass attenuation coefficient $\mu_{mass} = 0.0681$ cm$^2$/g by dividing for the 2.70 g/cm$^3$ aluminum density. The students compare the result that is obtained with the reference value of 0.0747 cm$^2$/g taken from the National Institute of Standards and Technology (NIST) database (https://physics.nist.gov). The students are stimulated to critically discuss the result, trying to justify the differences between the experimental and the reference value, due, for instance, to the presence of the stainless steel crystal housing or to the imperfect collimating effect of the hole in the lead shielding box. By checking the $\mu_{mass}$ value on the NIST database the students can appreciate from a numerical and graphical point of view its energy dependence and discuss the dominating interactions in different energy ranges.

Finally, the students are asked to make a further comparison between the experimental $\mu$ of aluminum and the NIST database value of $\mu$ for air ($9.52 \times 10^{-5}$ cm$^{-1}$). The discussion raised from this further comparison is functional in making the students master, through a joint and interdisciplinary approach, the concepts of the highly penetrating nature of the gamma radiation and of the dependence of the linear attenuation coefficient on the type of attenuating material.

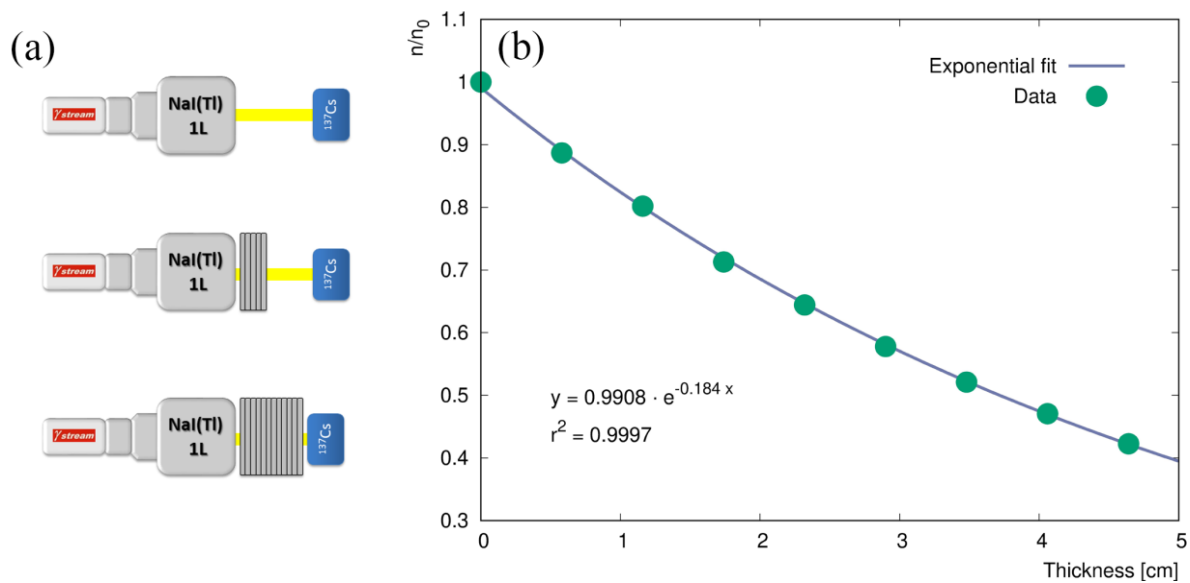

**Figure 6.** (**a**) A scheme of the different experimental configuration designed for the determination of the aluminum linear attenuation coefficient $\mu$. The students perform consecutive acquisitions by adding two aluminum layers each time, until the space between the detector and the $^{137}$Cs point-like source is completely filled. (**b**) The green points represent the experimental values of the ratio $\frac{n}{n_0}$ where $n_0$ and n are, respectively, the net count rates recorded in the $^{137}$Cs photopeak in the absence and in the presence of aluminum attenuating layers of given thickness. The blue curve is the exponential fit function from which the experimental $\mu$ of aluminum (0.184 cm$^{-1}$) is retrieved.

## 4. Outdoor Experiment

The outdoor experiment is dedicated to the design and the realization of in-situ γ-ray spectrometry measurements during which the potentialities of a client-server system are exploited. This hand-on experiment lets the students familiarize themselves with the factors affecting the different levels of terrestrial gamma radiation in the environment. Unlike a laboratory experience, in the outside environment, it is impossible to manage all of the parameters characterizing the experimental conditions. In the specific case of in-situ γ-ray spectroscopy, there are many variables that could interfere with the measurement, such as the presence of vegetation or buildings and the morphology of the area affecting the field of view of the spectrometer (Figure 2a) [15,16]. In addition, soil humidity has an attenuating effect on gamma radiation and sources having weak intensities need longer acquisition times. In order to compensate for these potential nuisances, a well-designed measurement procedure could help in minimizing the effects of outdoor factors. In this sense, students are stimulated to carefully adhere to the acquisition procedure, which comprises taking notes of all the relevant experimental conditions, especially in terms of their potential impact on the experimental outcomes. Finally, the results of the measurements are translated in a thematic map that can be visualized and then shared, together with the input data, on an open access Web-GIS platform via an interactive GUI.

The same detector (4" × 4" NaI(Tl)) described for the indoor experiment is employed for the outdoor survey, just arranged inside a backpack for portability. As the protagonists of the Ghostbusters movie, the students wear a backpack that makes them able to capture γ-ray spectra, which, like ghosts, are invisible to the eye. In outdoor campaigns, the potential of the instrument is fully exploited, as performing the measurement with just the use of an Android tablet simplifies and quickens the data taking operations.

*4.1. In-Situ Gamma-Ray Survey*

The in-situ survey is planned keeping in mind the spatial resolution of the desired information. The adopted strategy consists in choosing the sampling points in order to cover the surveyed area comprehensively for the different types of ground coverage, like asphalt, grass, or brick (Figure 7). A map of the measurement points located in and around the area is previously loaded in the Google Maps app of a smartphone. For each measurement point, the students perform a 180 s acquisition, take a picture of the area surrounding the detector, and compile the measurement sheet with GPS coordinates, type of ground coverage, and the measurement ID provided by the GammaTOUCH app. Knowing the detector field of view (Figure 2a), the students place the backpack at a sufficient distance from vertical structures (e.g., walls, trees) and avoiding standing close to the instrument during the acquisition. In this way, both the attenuation and the radiation emission effects of their bodies are minimized. During the data acquisition, the students are encouraged to make questions and assumptions, i.e., about the effects that a change in atmospheric conditions or soil humidity or about the type of ground coverage that would be the most or least abundant in natural radioactivity and to explain why, in order to support their views.

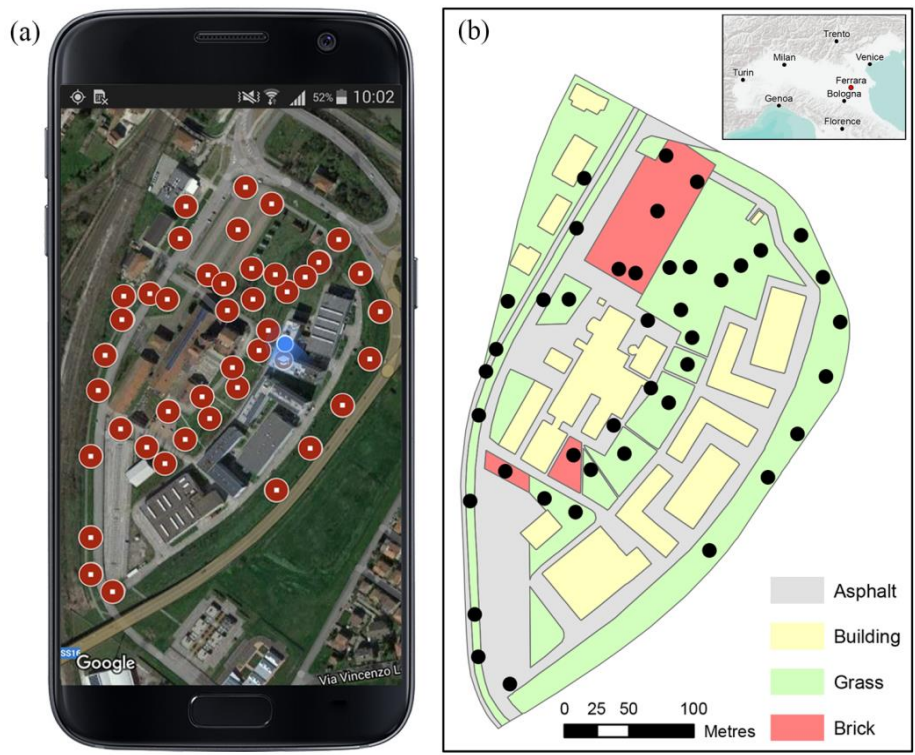

**Figure 7.** (**a**) Planned measurement points (red dots) reported in the Google Maps app. (**b**) Simplified map of the campus and the superimposed measurement points (black dots); the background colors indicate different types of ground coverage.

*4.2. From Counts to Radioelements Concentrations*

After the survey, the experimental results are organized in an Excel file, together with the information that is related to the data taking conditions. The Android app that is described in Section 4 is used for the energy calibration of the gamma spectra and for retrieving the total counts in the $^{40}$K, $^{214}$Bi, and $^{208}$Tl energy windows of interest (Figure 5) applying the Window Analysis Method [14]. The *K, U,* and *Th* concentrations (*C*) are determined by essentially applying the expression:

$$C = \frac{N}{S}, \quad (4)$$

where $N$ is the net count rate in the photopeak associated to the investigated element and $S$ is the sensitivity coefficient determined from the calibration measurements on the ground at natural sites [17].

The total specific activity $A$, measured in becquerel per kilogram (Bq/kg) due to the terrestrial radionuclides radiation is then determined as [14]:

$$A = 313 \cdot C_K + 12.35 \cdot C_U + 4.06 \cdot C_{Th}, \quad (5)$$

where $C_K$ is the potassium concentration in $10^{-2}$ g/g and $C_U$ and $C_{Th}$ are the uranium and thorium concentration in µg/g. More details about the analysis method as well as on the conversion from mass abundance to specific activity are provided in http://www.fe.infn.it/radioactivity/educational.

The results of the in-situ γ-ray measurements that were performed during the 2017 Maymester "Concepts in Nuclear and Radiation Engineering" are reported in Table 2 in terms of $K$, $U$, and $Th$ concentrations.

**Table 2.** Mean and standard deviation of the $K$, $U$, and $Th$ concentrations that were obtained from the in-situ γ-ray measurements distinguished according to the different ground coverage types. The measurements were performed during the 2017 Maymester.

| Ground Coverage | Number of Measurements | $K$ [$10^{-2}$ g/g] | $U$ [µg/g] | $Th$ [µg/g] |
|---|---|---|---|---|
| Brick | 7 | 0.82 ± 0.19 | 1.8 ± 0.5 | 4.1 ± 1.0 |
| Grass | 28 | 2.08 ± 0.32 | 1.7 ± 0.4 | 9.5 ± 1.8 |
| Asphalt | 7 | 1.20 ± 0.10 | 1.9 ± 0.4 | 5.1 ± 0.7 |

Adopting a Web-GIS platform, a kml file, a Google Earth supported format that is suitable for open access on-line publication, is created. The kml file reports the measurement points, each one assigned with the corresponding total specific activity and a picture of the acquisition location (Figure 8a). By easily inspecting the data reported in the kml file, the students are able to discuss the results, understand how radioactivity is spatially distributed, and how it relates to the ground coverage type.

*4.3. From Measurements to Map*

The measurements performed during the outdoor experiment are used to create the map of the natural radioactivity expressed in total specific activity in Bq/kg of the investigated area and a kml file for open access on-line publication (Figure 8b). The data spatialization is performed by adopting a multivariate geostatistical interpolator, the Collocated CoKriging [18] (CCoK). The CCoK is applied in order to predict the total specific activity, the under-sampled primary variable, using as constraint a secondary variable known in each location, i.e., the type of ground coverage. A continuous grid is created for the investigated area and a pseudo-random value is assigned to each type of coverage in order to obtain a normal distribution. The radiometric measurements are spatially conjoined to the coverage grid and a multivariate point dataset is obtained. The CCoK interpolation models are obtained by calculating experimental semi-variograms and experimental cross-semivariograms. Finally, the spatial interpolation is performed using a homogenous grid with a 10 m resolution. The students are encouraged to take under consideration the spatial resolution and distribution of γ-ray spectroscopy measurements.

The map of the natural radioactivity of the area investigated during the Maymester can be downloaded at http://www.fe.infn.it/radioactivity/educational (Mapping section), together with the kml files.

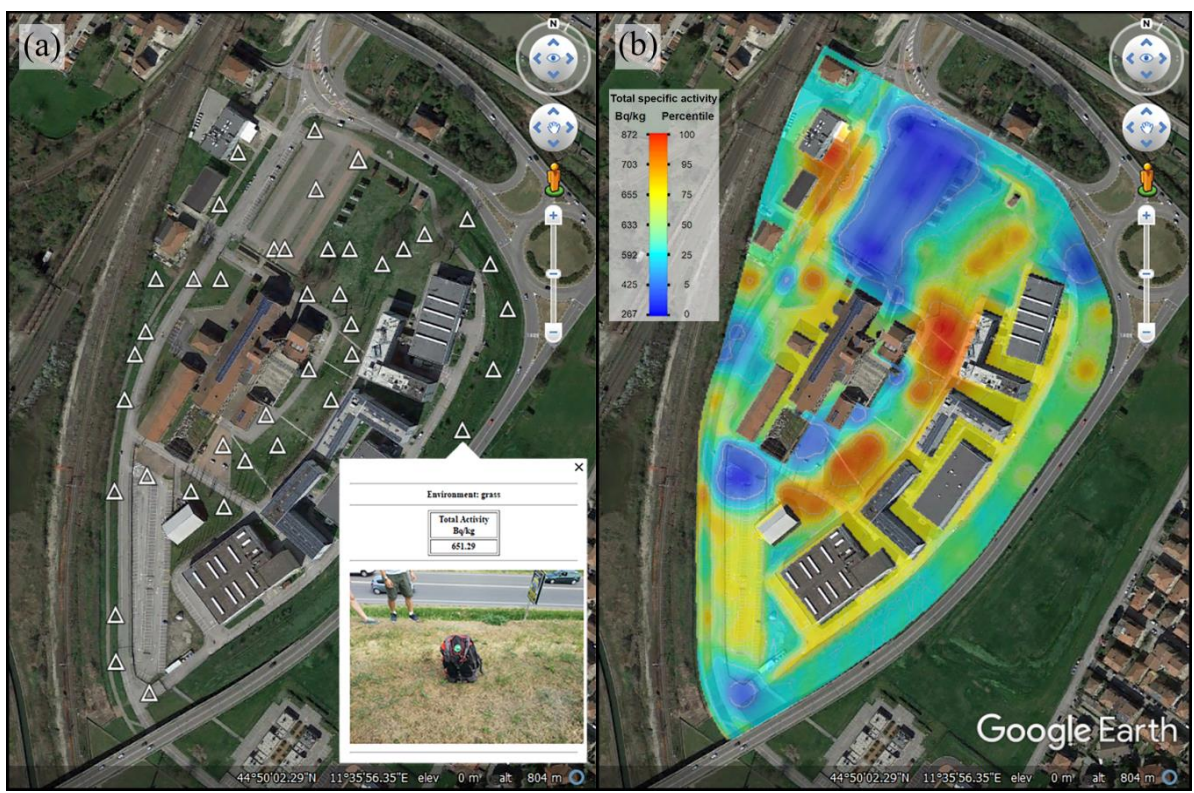

**Figure 8.** (**a**) Web-GIS tool for visualizing the measurement points (triangles) of natural radioactivity in the area investigated during the Maymester. By clicking on them a box containing the information gathered by the students shows up. (**b**) Map of the total specific activity in Bq/kg obtained from the spatial interpolation of the 42 in-situ γ-ray measurements.

## 5. Conclusions

The presented educational activities addressed to pre-collegiate students provide a successful example of the effectiveness of a mixed approach based on the use of engineering and computer science tools for conveying basic physics concepts that are related to the environmental radioactivity in-situ measurement. By reproducing these practical lectures, teachers are able to provide by means of a coherent multidisciplinary approach a method that is focused on a problem-solving attitude and consisting in analysis design, critical thinking, communication, and teamwork. The hands-on experiments are an opportunity for pre-collegiate students to break out of the traditional learning approach and to concretely tackle the challenges that a professional engineer could face.

By carrying out the indoor and the outdoor experiments, the students learnt how to perform a γ-ray spectroscopy measurement from the point of view of both hardware assembling and of software data taking and analysis. The gamma radiometric acquisition was discussed in all the relevant aspects, from the interpretation of the distinct features of the spectrum to the critical inspection of the experimental conditions that can potentially affect the outcome of the measurements. In the elaboration of the experimental data, the students are asked to work collaboratively in teams in order to extract from the acquired spectra quantitative information on the attenuating properties of a given material or to assess the radioactive content of different types of ground coverage.

During the indoor experiment, the students gather propaedeutic knowledge that is related to the interpretation of a γ-ray spectrum by identifying the distinct spectral features of the Compton continuum and of the photopeaks. Furthermore, the students quantitatively assess the high penetration nature of gamma radiation by estimating the linear attenuation coefficient of aluminum and by comparing it with the reference value and with that of air.

A total of 42 outdoor measurements were performed on brick, grass, and asphalt by using an Android app both for the managing of the experimental set up and for the retrieving of the net count rates in the natural radionuclide main photopeaks. The results that were obtained during the outdoor experiment in terms of total activity originating from *K*, *U,* and *Th* are visualized and published by means of a Google Earth kml file on an open access platform and they are synthetized in the natural radioactivity map of the investigated area.

Finally, the course assessment questionnaires were demonstrated to be an excellent source of feedback from the students regarding the educational content of the activities as well as the teaching methodologies.

**Author Contributions:** Conceptualization, M.A., M.B., S.L., V.S. and F.M.; Investigation, M.A., M.B., E.C., K.G.C.R. and F.M.; Methodology, M.A., M.B., E.C., K.G.C.R. and F.M.; Resources, M.A., M.B., E.C., K.G.C.R. and F.M.; Software, M.A., M.B. and E.C.; Writing – original draft, M.A., M.B., C.B., E.C., S.L., K.G.C.R., V.S. and F.M.; Writing – review & editing, M.A., M.B., S.L., K.G.C.R., A.S., V.S. and F.M..

**Funding:** This work was partially founded by the National Institute of Nuclear Physics (INFN) through the ITALian RADioactivity project (ITALRAD) and by the Theoretical Astroparticle Physics (TAsP) research network. The co-authors would like to acknowledge the support of the Geological and Seismic Survey of the Umbria Region (UMBRIARAD), of the University of Ferrara (Fondo di Ateneo per la Ricerca scientifica FAR 2017-2018).

**Acknowledgments:** The authors would like to thank the staff of Consorzio Futuro in Ricerca, CAEN Spa and GeoExplorer Impresa sociale s.r.l. for their support and Sandro Bardelli, Roberto Calabrese, Ivan Callegari, Giovanni Di Domenico, Adam Drescher, Adriano Duatti, Giovanni Fiorentini, Michele Montuschi, Andrea Motti, Ferruccio Petrucci, Barbara Ricci, Raffaele Tripiccione and Gerti Xhixha. We are indebted to the University of Texas Cockrell School of Engineering and Ellen Aoki, Senior Academic Program Coordinator, and the Study Abroad Program in the International Office for organizing the one month course in 2016, 2017 and 2018.

## References

1. Siegel, P. Gamma spectroscopy of environmental samples. *American Journal of Physics* **2013**, *81*, 381-388.
2. Pilakouta, M.; Savidou, A.; Vasileiadou, S. A laboratory activity for teaching natural radioactivity. *European Journal of Physics* **2016**, *38*, 015801.
3. Prather, E. Students' beliefs about the role of atoms in radioactive decay and half-life. *Journal of Geoscience Education* **2005**, *53*, 345-354.
4. Neumann, S.; Hopf, M. Students' conceptions about 'radiation': Results from an explorative interview study of 9th grade students. *Journal of Science Education and Technology* **2012**, *21*, 826-834.
5. Anjos, R.; Okuno, E.; Gomes, P.; Veiga, R.; Estellita, L.; Mangia, L.; Uzeda, D.; Soares, T.; Facure, A.; Brage, J. Radioecology teaching: evaluation of the background radiation levels from areas with high concentrations of radionuclides in soil. *European Journal of Physics* **2003**, *25*, 133.
6. Anjos, R. Radioecology teaching: response to a nuclear or radiological emergency. *European journal of physics* **2006**, *27*, 243.
7. Peralta, L. Measuring the activity of a radioactive source in the classroom. *European journal of physics* **2004**, *25*, 211.
8. UNSCEAR. *Sources and effects of ionizing radiation. Volume I: Sources Report to the General Assembly, with Scientific Annexes*; United Nations Publications: 2000.
9. Kocher, D.C. *Radioactive decay data tables*; Oak Ridge National Lab.: 1981.
10. Tsoulfanidis, N.; Landsberger, S. *Measurement and detection of radiation*, Fourth ed.; Group, T.F., Ed. CRC press: Boca Raton, 2015.
11. Baldoncini, M.; Albéri, M.; Bottardi, C.; Chiarelli, E.; Raptis, K.G.C.; Strati, V.; Mantovani, F. Investigating the potentialities of Monte Carlo simulation for assessing soil water content via proximal gamma-ray spectroscopy. *Journal of Environmental Radioactivity* **2018**, *192*, 105-116.
12. Kucuk, N.; Tumsavas, Z.; Cakir, M. Determining photon energy absorption parameters for different soil samples. *Journal of radiation research* **2012**, *54*, 578-586.
13. Adamides, E.; Koutroubas, S.; Moshonas, N.; Yiasemides, K. Gamma-ray attenuation measurements as a laboratory experiment: some remarks. *Physics Education* **2011**, *46*, 398.
14. IAEA. Guidelines for radioelement mapping using gamma ray spectrometry data. *International Atomic Energy Agency, Vienna. Technical Report Series No. 323* **2003**.

15. Xhixha, M.K.; Albèri, M.; Baldoncini, M.; Bezzon, G.; Buso, G.; Callegari, I.; Casini, L.; Cuccuru, S.; Fiorentini, G.; Guastaldi, E. Uranium distribution in the Variscan Basement of Northeastern Sardinia. *Journal of Maps* **2016**, *12*, 1029-1036.

16. Baldoncini, M.; Albèri, M.; Bottardi, C.; Chiarelli, E.; Raptis, K.G.C.; Strati, V.; Mantovani, F. Biomass water content effect in soil water content assessment via proximal gamma-ray spectroscopy. *Geoderma* **2019**, *335*, 69-77, doi:https://doi.org/10.1016/j.geoderma.2018.08.012.

17. Caciolli, A.; Baldoncini, M.; Bezzon, G.; Broggini, C.; Buso, G.; Callegari, I.; Colonna, T.; Fiorentini, G.; Guastaldi, E.; Mantovani, F. A new FSA approach for in situ γ ray spectroscopy. *Science of the Total Environment* **2012**, *414*, 639-645.

18. Guastaldi, E.; Baldoncini, M.; Bezzon, G.; Broggini, C.; Buso, G.; Caciolli, A.; Carmignani, L.; Callegari, I.; Colonna, T.; Dule, K. A multivariate spatial interpolation of airborne γ-ray data using the geological constraints. *Remote sensing of environment* **2013**, *137*, 1-11.